\documentclass[twocolmun,aps,11pt]{revtex4}
\usepackage{epsfig}
\usepackage{amsmath}
\usepackage{bm}
\usepackage{times}
\usepackage{graphicx}
\usepackage{color}
\def\beq{\begin{equation}}
\def\eeq{\end{equation}}
\def\bea{\begin{eqnarray}}
\def\eea{\end{eqnarray}}

	\newcommand{\abs}[1]{ \mathopen{}\left| {#1}\right| }
	\DeclareMathOperator{\real}{Re}
	\DeclareMathOperator{\imag}{Im}
	
\begin{document}
\title{THE RIGHT SIDE OF TEV SCALE SPONTANEOUS R-PARITY VIOLATION}
\bigskip
\author{Lisa L.~Everett}
\author{Pavel Fileviez P{\'e}rez}
\author{Sogee Spinner}
\address{
Department of Physics, University of Wisconsin, Madison, WI 53706, USA}
\date{\today}

\begin{abstract}
We study a simple extension of the minimal supersymmetric Standard Model in which 
the Abelian sector of the theory consists of $B-L$ and right-handed isospin. In the minimal 
model this Abelian gauge structure is broken to the standard model hypercharge gauge group 
by non-vanishing vacuum expectation values of the right-handed sneutrinos, resulting 
in spontaneous R-parity violation. This theory can emerge as a low energy effective theory 
of a left-right symmetric theory realized at a high scale. We determine the mass spectrum 
of the theory, and discuss the generation of neutrino masses and R-parity  violating interactions. The possibility of distinguishing between R-parity violating models with a gauged $U(1)_{B-L}$ broken at the TeV scale at the Large Hadron Collider is discussed.
\end{abstract}

\maketitle

\section{Introduction}
\label{section1}
In models of physics beyond the Standard Model (SM) with low energy supersymmetry (SUSY), such as the minimal supersymmetric standard model (MSSM) and its extensions,  a conserved parity symmetry is usually added to prevent rapid proton decay from dimension four operators. The most common symmetry of this type is R-parity, which is defined as $R=(-1)^{3(B-L) + 2 S}= (-1)^{2S} M$, in which $M$ is called matter parity and $B$, $L$ and $S$ stand for baryon number, lepton number, and spin, respectively. The inclusion of a conserved R-parity not only ensures the stability of the proton, but also allows for a stable lightest superpartner (LSP) that is a good candidate for the dark matter of the universe. Hence, understanding the origin of R-parity, and whether it is an exact symmetry or is broken spontaneously at some scale, is of key importance for theories with low energy supersymmetry (see e.g.~\cite{Haber,newrevs} for general reviews of supersymmetry and~\cite{hallsuzuki,gen,mohapatra,sneutrino,Martin,susyLR,susyso10} for discussions of R-parity violation in supersymmetric models). 

To understand if R-parity is conserved or broken at energies relevant for tests of supersymmetry at the Large Hadron Collider (LHC), one can see from the definition of R-parity that it is useful to investigate scenarios in which $B-L$ is an exact symmetry of the theory at some scale (note that if  $B-L$ is conserved, matter parity is also conserved). Such scenarios are also well-motivated from the theoretical point of view in that $B-L$ emerges naturally in the context of many grand unified theories.  However, for minimal models in which $B-L$ is a local symmetry, if the matter content of the theory includes only the MSSM states (including right-handed neutrinos) such that no additional fields are introduced by hand, the inevitable consequence is that R-parity is spontaneously broken.  

Scenarios with a spontaneously broken gauged $B-L$ symmetry have the attractive feature that the R-parity breaking in the low energy theory is obtained in a controlled manner.  In such scenarios, rapid proton decay can be avoided because baryon number can remain as a symmetry that is conserved at the perturbative level. Though the LSP neutralino dark matter framework is not applicable once R-parity is broken, it is well known that the gravitino can be a viable dark matter candidate.  These models not only have the distinctive collider signatures associated with R-parity violation, but also if the gauged $B-L$ symmetry is broken at the TeV scale, the gauge structure of this sector can be probed in detail at the LHC.

To this end, two of the authors have studied extensions of the MSSM in which $B-L$ is part of the local gauge symmetry, but is broken at TeV energies by non-vanishing sneutrino vacuum expectation values.  These models have included both left-right symmetric models~\cite{Sogee1} and simple $U(1)_{B-L}$ extensions of the MSSM \cite{Sogee2,Sogee3}.  Of these scenarios, the left-right symmetric model of ~\cite{Sogee1} is particularly minimal and economical in that it contains the simplest possible Higgs sector needed for the breaking of the left-right symmetry. However, for this scenario to be viable, the soft-breaking terms must not respect the discrete left-right parity symmetry, and hence it is not possible to simultaneously understand the origin of parity violation and R-parity violation in this context. It is also difficult to satisfy all of the constraints arising from flavor violation due to additional light Higgses in the theory.  In contrast, the simple  $U(1)_{B-L}$ extensions of the MSSM studied 
in~\cite{Sogee2,Sogee3} are anomaly-free $U(1)^\prime$ theories with distinctive collider phenomenology due to R-parity breaking and the TeV scale $Z^\prime$ gauge boson associated with the gauged $B-L$ symmetry. For phenomenological and cosmological aspects of the R-parity violating interactions see~\cite{R-parity}. 

In this work, we investigate spontaneous R-parity violation within a model which naturally includes the best features 
of each of these scenarios. The model is a simple extension of the MSSM in which the Abelian sector of the theory 
consists of $U(1)_{B-L}$ and $U(1)_{I^R_3}$, the third component of right-handed weak isospin (the $SU(2)_R$ of left-right models). This theory is thus a simple $U(1)_{B-L}$ extension of the MSSM, but it can also be obtained from a left-right symmetric theory that is realized at some high scale when the $SU(2)_R$ is broken to $U(1)_{I^R_3}$ by an $SU(2)$ triplet Higgs with 
zero $B-L$ charge. We determine the full spectrum of the theory, and discuss the generation of neutrino masses and 
the properties of the Higgs sector. We also discuss ways to distinguish between this class of models via their $Z^\prime$ phenomenology at the LHC.  
 
This work is organized as follows. In Section~\ref{section2}, we present our model and discuss spontaneous R-parity violation. In Section III, we compute the full mass spectrum (including neutrino masses), and outline the details of the R-parity violating interactions. We discuss the methods for distinguishing  between these different models for spontaneous R-parity violation  in Section IV. Finally, we summarize our main results and conclude in Section V.
\section{Spontaneous R-parity Violation in a Simple Extension of the MSSM}
\label{section2}
We consider an anomaly-free extension of the MSSM in which the electroweak gauge sector consists of 
$  \ SU(2)_L \bigotimes U(1)_{I_3^R} \bigotimes U(1)_{B-L} $, in which $U(1)_{I_3^R}$ is defined to be the 
third component of right-handed isospin ($SU(2)_R$).  More precisely, the quark and lepton superfields 
have the electroweak gauge charges:
\begin{eqnarray}
\hat{Q}^T &=& \left(\hat{U}, \hat{D}\right) \sim (2,0,1/3), 
\;\;\;\; \hat{U}^C \sim (1, -1/2, -1/3),\nonumber \\
\hat{D}^C& \sim & (1,1/2,-1/3),
 \qquad \hat{L}^T = \left( \hat{N}, \hat{E} \right) \sim (2,0,-1),\nonumber \\
 \hat{E}^C &\sim & (1,1/2,1), \qquad \qquad \hat{N}^C \sim (1,-1/2,1).
 \label{matt}
 \end{eqnarray}
The MSSM Higgses transform as 
\begin{equation}
\label{higgs}
\hat{H}_u \sim (2,1/2,0),\qquad \hat{H}_d \sim (2,-1/2,0).
\end{equation}
This minimal set of superfields as described above is the full matter content within our model. Therefore, in order to break $U(1)_{I_3^R} \bigotimes U(1)_{B-L} \rightarrow U(1)_Y$, the sneutrinos must acquire non-zero vacuum expectation values, resulting in spontaneous R-parity violation (for previous discussions of sneutrino vacuum expectation values, see e.g.~\cite{gen,sneutrino}). Hence, within the minimal model (in terms of particle content) with this Abelian gauge structure,  
R-parity should be broken spontaneously.

Before exploring the details of the gauge symmetry breaking and resulting R-parity violation in this model, we pause to comment on the connection of this scenario to left-right symmetric models (for previous discussions, see \cite{susyLR}).  The gauge structure described above can be obtained as a low energy limit of a left-right symmetric theory based on the gauge group $SU(3)_C \bigotimes SU(2)_L \bigotimes SU(2)_R \bigotimes U(1)_{B-L}$.  The $SU(2)_R$ symmetry is broken at a high scale by a triplet  $\Sigma^C \sim (1,1,3,0)$ (generically present in this class of models as part of a pair of triplets, together with $\Sigma \sim (1,3,1,0)$) to the Abelian subgroup $U(1)_{I^R_3}$ \cite{susyLR}.  
The origin of the parity violating interactions present in the MSSM (or in the SM) can then be understood once the left-right symmetry is spontaneously broken at the high scale.

Indeed, spontaneous R-parity violation has been previously explored within a minimal fully left-right symmetric model in which the Higgs sector is composed only of Higgs bi-doublets, $\Phi_i \sim (2,2,0)$ ~\cite{Sogee1}.   As discussed in~\cite{Sogee1}, the scenario can be consistent only if soft supersymmetry breaking terms do not respect the discrete left-right symmetry ({\it i.e.}, if parity is broken in the soft breaking sector). To avoid strong mixing between the leptons and the charginos/neutralinos, the left-right symmetry should be broken at the TeV scale. However, in this case the several light Higgses present in the minimal version of the theory can result in unacceptably large flavor violation.  In contrast, within the $SU(2)_L \bigotimes U(1)_{I_3^R} \bigotimes U(1)_{B-L}$ scenario described in this paper, the scales of R-parity violation and left-right symmetry breaking ({\it i.e.}, parity violation) are logically separated, and issues of flavor violation can be safely avoided. 

Given the particle content in Eqs.~(\ref{matt})--(\ref{higgs}), the superpotential is given by      
\begin{equation}
{\cal W}_{R}={\cal W}_{MSSM} \ + Y_\nu \ \hat{L}  \hat{H}_u \hat{N}^C,
\end{equation}
in which the MSSM superpotential takes the form
\begin{eqnarray}
{\cal W}_{MSSM} &=& Y_u \ \hat{Q}  \hat{H}_u \hat{U}^C + Y_d \ \hat{Q} \hat{H}_d \hat{D}^C \ + \ Y_e \ \hat{L} \hat{H}_d \hat{E}^C + \mu \ \hat{H}_u  \hat{H}_d.
\end{eqnarray}
In the above, we have suppressed flavor, gauge, and Lorentz indices.  We do not adhere to any specific model of SUSY breaking, and hence the soft breaking terms are:
\begin{eqnarray}
	\nonumber
	V_{soft} & = & M_{\tilde N^C}^2 \vert \tilde{N}^C\vert^2 \ + \ M_{\tilde L}^2 \ \vert \tilde L\vert^2 + M_{\tilde E^C}^2 \ \vert \tilde E^C \vert^2 
	\ + \ m_{H_u}^2 \abs{H_u}^2 + m_{H_d}^2 \abs{H_d}^2 \nonumber \\ &+& 
	\ \left(  A_\nu \ \tilde{L} H_u \ \tilde{N}^C +  B \mu \ H_u  \ H_d + \frac{1}{2} M_{BL} \tilde{B^{'}} \tilde{B^{'}}  
	\right. \ + \
	  	\left. \frac{1}{2} M_R  \ \tilde{W}_R^0 \tilde{W}_R^0  \  + \mathrm{h.c.} \right)  \ +   V_{soft}^{MSSM}.
\label{soft}
\end{eqnarray}
In Eq.~(\ref{soft}), $M_{BL}$ and $M_R$ denote the gaugino masses for the $U(1)_{B-L}$ and $U(1)_{I_3^R}$ gauginos, 
and $V_{soft}^{MSSM}$ represents terms that are not relevant for our discussion of spontaneous 
R-parity violation in this model ({\it i.e.}, the remaining gaugino masses, soft mass-squared parameters, and trilinear couplings of the MSSM).

We now discuss  gauge symmetry breaking and its connection to R-parity violation in this model.
It is straightforward to compute the different contributions to the potential once one generation of left-handed and right-handed sneutrinos ($\tilde{\nu}$ and $\tilde{\nu}^C$), and the Higgses ($H_{u,d}$), acquire vacuum expectation values (VEVs) $\langle \tilde{\nu} \rangle =v_L/\sqrt{2}$, $\langle \tilde{\nu}^C \rangle = v_R/\sqrt{2}$, and $\langle H^0_{u,d} \rangle = v_{u,d}/\sqrt{2}$, respectively. 
These contributions read as
\begin{eqnarray}
	\left<V_F \right>  &= &
		\frac{1}{4} \left(Y_\nu \right)^2
		\left(
			v_R^2 v_u^2 + v_R^2 v_L^2 + v_L^2 v_u^2
		\right) 
		\ + \  \ \frac{1}{2} \mu^2
		\left(
			v_u^2 + v_d^2
		\right)
		 -  \frac{1}{\sqrt{2}} Y_\nu \ \mu \ v_d  v_L v_R,
	\\
	\left<V_D \right> &=&
		\frac{1}{32}
		\left[
			g_2^2
			\left(
				 v_u^2 -v_d^2 - v_L^2
			\right)^2
			+ g_R^2
			\left(
				v_d^2 + v_R^2 - v_u^2
			\right)^2 \right. 
	 \ + \
			\left. g_{BL}^2
			\left(
				v_R^2 - v_L^2
			\right)^2
		\right],
	\\
	\left<V_{soft} \right>  &=&
		\frac{1}{2} \left (M_{\tilde L}^2 v_L^2 + M_{\tilde N^c}^2 v_R^2 + m_{H_u}^2 v_u^2 +  m_{H_d}^2 v_d^2 \right ) 
		\ + \ \frac{1}{2 \sqrt{2}} (A_\nu + A^\dagger_\nu) \ v_R v_L v_u
		-  \text{Re} \left( B \mu \right) \ v_u v_d,
		\nonumber \\
\end{eqnarray}
in which $g_R$, $g_2$ and $g_{BL}$ are the gauge couplings for $U(1)_{I_{3}^R}$, 
$SU(2)_L$ and $U(1)_{B-L}$, respectively. Minimizing the potential 
in the limit $v_R, \ v_u, \ v_d \gg v_L$ yields:
\begin{align}
\label{vRvev}
	v_R & = \sqrt
				{
					\frac{-8 M_{\tilde{N}^C}^2 + g_R^2 \left(v_u^2 - v_d^2\right)}
					{g_R^2 + g_{BL}^2}
				},
	\\
	\label{vLvev}
	v_L & = \frac{v_R B_\nu}
			{
				M_{\tilde L}^2 - \frac{1}{8} g_{BL}^2 v_R^2 - \frac{1}{8} g_2^2 \left(v_u^2 - v_d^2\right)
			},
	\\
	\nonumber
	\\
	\label{mu}
	\mu^2  = - \frac{1}{8} \left(g_2^2 + g_R^2\right) \left(v_u^2 + v_d^2\right)
			& + \frac{M_{H_u}^2 \tan^2 \beta  - M_{H_d}^2}{1-\tan^2 \beta},
	  \  \  \ {\rm and} \  \  \
	B\mu  =  2 \mu^2 + m_{H_d}^2 + m_{H_u}^2,
\end{align}
in which 
\begin{align}
	\label{MHu}
	M_{H_u}^2 & = m_{H_u}^2 - \frac{1}{8} g_R^2 v_R^2,
	\  \  \
	M_{H_d}^2  = m_{H_d}^2 + \frac{1}{8} g_R^2 v_R^2,
	\  \  \ {\rm and} \ \ \
B_\nu  = \frac{1}{\sqrt{2}} \left(Y_\nu \mu v_d - A_\nu v_u \right).
\end{align}
From Eqs.~(\ref{vRvev})--(\ref{vLvev}), we see that while $v_L\neq 0$ if $v_R\neq 0$, $v_L$ is generically 
suppressed compared to $v_R$.  We also see from Eq.~(\ref{vRvev}) that the desired symmetry breaking pattern requires negative soft breaking terms for the right-handed sneutrinos, $M_{\tilde{N}^C}^2 < 0$.  Such negative soft mass-squares can in principle be achieved via a radiative mechanism in the context  of a high energy theory or in some gauge mediation mechanism.   However, it is also worth noting that this requirement can be circumvented in the presence of a Fayet-Iliopoulos (FI) terms for the Abelian symmetries, as shown in Appendix B. For our purposes for this paper, we will restrict ourselves to the case without such FI terms and simply assume the presence of the required soft breaking terms at low energies.  The remaining minimization condition, 
Eq.~(\ref{mu}), is similar to the MSSM case except that Eq.~(\ref{mu}) now has extra contributions 
from the $I_{3R}$ $D$-terms.  As indicated by Eq.~(\ref{MHu}), these contributions are consistent 
with having a negative mass squared for the up-type Higgs. 
\section{Bilinear RpV interactions and Mass Spectrum}
\label{section3}
Once the sneutrinos acquire VEVs, R-parity is spontaneously broken and the low energy theory is similar to the bilinear R-parity violating MSSM.  The traditional bilinear R-parity violating term is given by $Y_\nu v_R L \ \tilde H_u$, with coefficient suppressed by neutrino mass parameters.  The kinetic term of the right-handed neutrino leads to a new R-parity violating mixing with the gaugino
$g_R v_R \nu^C \tilde B'$, which is not suppressed by neutrino masses, but does not directly lead to dangerous low energy observables.  Further bilinears can be derived in a similar fashion and will be proportional to $v_L$ and therefore suppressed by neutrino mass parameters.  Effective trilinear terms appear once the neutralinos are integrated out and will be doubly suppressed and proportional to the ratio of neutrino mass parameters to neutralino masses.  It is important to emphasize once more that the baryon violating $\lambda{''}$ will not be generated and therefore neither will dimension four proton decay operators.

When considering the gauge sector,  we will work in the limit that $v_L \to 0$ for simplicity.
Defining $v^2 \equiv v_u^2 + v_d^2$ as usual, the neutral gauge boson mass matrix in the $(B', W_R^0, W_L^0)$  basis is given by 
\begin{equation}
	\label{Z.Mass.Matrix}
	\mathcal{M}_Z^2 =
	\begin{pmatrix}
		\frac{1}{4}g_{BL}^2 v_R^2
		&
		-\frac{1}{4} g_{BL} g_R v_R^2
		&
		0
		\\
		-\frac{1}{4} g_{BL} g_R v_R^2
		&
		\frac{1}{4} g_R^2 \left(v_R^2 + v^2\right)
		&
		-\frac{1}{4} g_2 g_R v^2
		\\
		0
		&
		-\frac{1}{4} g_2 g_R v^2
		&
		\frac{1}{4} g_2^2 v^2
	\end{pmatrix},
\end{equation}
\\
in which ${\rm{Tr}} \ {\mathcal M}_Z^2 = M_{Z^{'}}^2 \ + \ M_{Z}^2 = (g_{BL}^2 + g_R^2) v_R^2/4  \ + \ (g_R^2 + g_2^2) v^2/4 $.  The neutral gauge boson masses are given by
\begin{align}
	M_{Z,Z'}^2
		& = \frac{1}{8}
		\left[
			\left(g_2^2 + g_R^2\right) v^2
			+ \left(g_{BL}^2 + g_R^2\right) v_R^2
		\right. 
		\left.
		\mp 
		\sqrt
		{
			\left(
				\left(g_2^2 + g_R^2\right) v^2
				- \left(g_{BL}^2 + g_R^2\right) v_R^2
			\right)^2
			+4 g_R^4 v^2 v_R^2
		}\
		\right]
\end{align}
and, of course, the massless photon. Expanding in $\epsilon^2 \equiv v^2/v_R^2$ results in:
\begin{eqnarray}
	\label{MzMass}
	M_{Z}^2 &\approx&
		\frac{1}{4} \left(\frac{g_R^2 g_{BL}^2}{g_R^2 + g_{BL}^2} + g_2^2\right) v^2
		\left(
			1 - \frac{g_R^4 \ \epsilon^2}{\left(g_R^2 + g_{BL}^2\right)^2 }
		\right),
	\\
	M_{Z'}^2 & \approx &
		\frac{1}{4} \left(g_R^2 + g_{BL}^2\right) v_R^2 
		\left(
			1 + \frac{g_R^4 \ \epsilon^2}{\left(g_R^2 + g_{BL}^2\right)^2 }
		\right),
\end{eqnarray}
where one can define $g_1^2 \equiv g_R^2 g_{BL}^2/(g_R^2 + g_{BL}^2)$ as the analogue of the
standard model hypercharge gauge coupling.  The $\rho$ parameter constrains $\epsilon^2
\lesssim 10^{-3}$.

We proceed in the basis
\begin{eqnarray}
	A & = &
		-\cos \theta_W \sin \theta_R B'
		+\cos{\theta_W} \cos \theta_R {W_R^0}
		\ + \ \sin{\theta_W} {W_L^0},
	\\
	Z_L & =&
		\sin \theta_W \sin \theta_R B'
		-\sin \theta_W \cos \theta_R {W_R^0}
		\ + \ \cos \theta_W {W_L^0},
	\\
	Z_R & =&
		\cos \theta_R B'
		+\sin \theta_R {W_R^0},
\end{eqnarray}
where, $\theta_W$ is the weak mixing angle and $\theta_R$ is the equivalent mixing angle of the
right-handed sector with $\tan{\theta_R} = -g_{R}/g_{BL}$. In this basis, the massless photon ($A$) is
a mass eigenstate and is projected out.  The remaining $2\times 2$ submatrix mixes $Z_L$ and $Z_R$.  This mixing angle $\xi$ can be parameterized as
\begin{equation}
	\tan{2 \xi} = 2 \frac{M_{Z_L Z_R}^2}{M_{Z_R}^2 - M_{Z_L}^2},
\end{equation}
where
\begin{align}
M_{Z_L Z_R}^2 & = \frac{1}{4} \ \frac{g_R^2}{\sqrt{g_{BL}^2 + g_R^2}} \ v^2 \sqrt{g_1^2+ g_2^2},
	\ \ \
	M_{Z_L}^2  = \frac{1}{4} \ (g_1^2 + g_2^2) \ v^2, 
\end{align}
and
\begin{align}
	M_{Z_R}^2 & = \frac{1}{4} \left(g_R^2 + g_{BL}^2\right) v_R^2.
\end{align}
Keeping terms to order $\epsilon^2$ yields:
\begin{equation}
	\xi \approx
		\frac{g_R^2 \sqrt{g_1^2 + g_2^2}}
		{\left(g_{BL}^2 + g_R^2\right)^{\frac{3}{2}}} \ \epsilon^2,
\end{equation}
with mass eigenstates
\begin{align}
	Z & = Z_L \cos \xi + Z_R \sin \xi,
	\\
	Z' & = Z_R \cos \xi - Z_L \sin \xi.
\end{align}
Experimentally, the $Z-Z^\prime$ mixing angle $\xi$ must be smaller than $\sim 10^{-3}$, which places a similar constraint on $\epsilon$ as the $\rho$ parameter.  To zeroth order in $\xi$, the couplings of the $Z'$ gauge boson to the fermions are
\begin{equation}
	g_{Z' \bar f f} = 
	\sqrt{g_R^2 + g_{BL}^2} \
		\left[
			\cos \theta_R^2 \ Y(f)
			- I_{3R}(f)
		\right],
\end{equation}
where $Y(f)$ and $I_{3R}(f)$ denote the hypercharge and the third component of the right-handed isospin.

Once R-parity is spontaneously broken, mixing between the neutralinos and the neutrinos is induced in the low energy theory.  It is well known, and has been extensively been discussed in the literature, that this mixing contributes to neutrino masses (for previous discussions, see e.g~\cite{hallsuzuki,gen,neutrinonew,neutrinoLR}).  In the $\left(\nu, \ \nu^c, \ \tilde B^{'}, \ \tilde W_R^0,\ \tilde W_L^0, \ \tilde H_d^0, \ \tilde H_u^0 \right)$ basis, 
the mass matrix is given by
\begin{widetext}
\begin{equation}
	{\cal M}_{N} =
	\begin{pmatrix}
			0
		&
			\frac{1}{\sqrt{2}} Y_\nu v_u
		&
			-\frac{1}{2} g_{BL} v_L
		&
			0
		&
			\frac{1}{2} g_2 v_L
		&
			0
		&
			\frac{1}{\sqrt{2}} Y_\nu v_R
	\\
			\frac{1}{\sqrt{2}} Y_\nu v_u
		&
			0
		&
			\frac{1}{2} g_{BL} v_R
		&
			\frac{1}{2} g_{R} v_R
		&
			0
		&
			0
		&
			\frac{1}{\sqrt{2}} Y_\nu v_L
	\\
			-\frac{1}{2} g_{BL} v_L
		&
			\frac{1}{2} g_{BL} v_R
		&
			M_{BL}
		&
			0
		&
			0
		&
			0
		&
			0
	\\
			0
		&
			\frac{1}{2} g_R v_R
		&
			0
		&
			M_R
		&
			0
		&
			\frac{1}{2} g_R v_d
		&
			-\frac{1}{2} g_R v_u
	\\
			\frac{1}{2} g_2 v_L
		&
			0
		&
			0
		&
			0
		&
			M_2
		&
			\frac{1}{2} g_2 v_d
		&
			-\frac{1}{2} g_2 v_u
	\\
			0
		&
			0
		&
			0
		&
			\frac{1}{2} g_R v_d
		&
			\frac{1}{2} g_2 v_d
		&
			0
		&
			-\mu
	\\
			\frac{1}{\sqrt{2}} Y_\nu v_R
		&
			\frac{1}{\sqrt{2}} Y_\nu v_L
		&
			0
		&
			-\frac{1}{2} g_R v_u
		&
			-\frac{1}{2} g_2 v_u
		&
			-\mu
		&
			0
	\end{pmatrix}.
\label{neutralino}
\end{equation}
\end{widetext}
Integrating out the neutralinos generates masses for both the right-handed neutrinos and the light neutrinos.  The right-handed neutrinos can then also be integrated out, resulting in further contributions to the light neutrino masses. The light neutrino masses are therefore generated through a double-seesaw mechanism, which includes both R-parity violating and Type I seesaw contributions.  For phenomenologically acceptable neutrino masses, the neutrino Yukawa coupling $Y_\nu$ must be of the order of $\sim 10^{-6}$, even for the third generation.  It is important to note that within left-right models, it is a challenge to obtain such a small neutrino Yukawa coupling and simultaneously generate  appropriate charged lepton masses, although it is not impossible.  Such models require two bi-doublets, both of which contains an up-type and down-type Higgs doublet and each with a different lepton Yukawa coupling.  The Dirac masses of the charged leptons and the neutrinos are then different linear combinations of the product of the lepton Yukawa couplings and the appropriate VEVs of these bi-doublet fields and it is possible for one of these linear combinations to be quite small, while keeping the other large enough for the charged lepton masses. In the context of this work, in which we focus primarily on the effective TeV-scale theory involving the gauged $B-L$ and $I_3^R$ rather than its parent left-right theory, we will not address this issue further.

Therefore, in the limit that $Y_\nu$ is neglected for simplicity,  the right-handed neutrino masses are given by $M_{\nu^C} \approx g_R^2 v_R^2 / M_R \ + \ g_{BL}^2 v_R^2 / M_{BL}$. Hence for neutralino masses below a TeV, the right-handed neutrino masses are on the order of 100 GeV.  Recall that in the case of the Type I seesaw~\cite{TypeI}, the light neutrino masses are given by $M_\nu = M_\nu^D M_{\nu^C}^{-1} (M_{\nu}^D)^T$.  In this case, $M_{\nu}^D$  has two different contributions: the first is governed by $Y_{\nu}$, and the second results from integrating out the neutralinos.  Aside from the light neutrinos, there will be five electroweak mass neutralinos.  Notice from the mass matrix above that substantial mixing exists between the right-handed neutrinos and the new Abelian gauginos and is of order TeV.

In the scalar sector of the theory, the sleptons and the Higgs scalars also mix as a result of R-parity violation.  Defining the $\sqrt{2} \imag \left(\tilde \nu, \tilde \nu^c, H_d^0, H_u^0 \right)$ basis for 
the CP-odd neutral scalars, the $\sqrt{2} \real \left(\tilde \nu, \tilde \nu^c, H_d^0, H_u^0 \right)$ basis 
for the CP-even neutral scalars,  and the $\left(\tilde e^*, \tilde e^c, H_d^{-*}, H_u^+ \right)$ basis for the charged scalars, the scalar mass-squared matrices are given by Eq.~(\ref{CP-odd}), Eq.~(\ref{CP-even}) and Eq.~(\ref{Charged}),
respectively. The mass matrix of the CP-odd neutral Higgses is given by
\begin{widetext}
\begin{equation}
\label{CP-odd}
{\cal M}_{P}^2
	=
	\begin{pmatrix}
	  	\frac{v_R}{v_L} B_\nu
	&
		B_\nu
	&
		-\frac{1}{\sqrt{2}} Y_\nu \mu \ v_R
	&
		-\frac{1}{\sqrt{2}} A_\nu v_R
\\
		B_\nu
	&
		\frac{v_L}{v_R} B_\nu
	&
		-\frac{1}{\sqrt{2}} Y_\nu \mu \ v_L
	&
		-\frac{1}{\sqrt{2}} A_\nu v_L
\\
		-\frac{1}{\sqrt{2}} Y_\nu \mu \ v_R
	&
		-\frac{1}{\sqrt{2}} Y_\nu \mu \ v_L
	&
		\frac{v_u}{v_d} B\mu \ + \frac{Y_\nu \mu \ v_L v_R}{\sqrt{2} \ v_d}
	&
		B\mu
\\
		-\frac{1}{\sqrt{2}} A_\nu v_R
	&
		-\frac{1}{\sqrt{2}} A_\nu v_L
	&
		B\mu
	&
		\frac{v_d}{v_u} B\mu \ - A_\nu \frac{v_L v_R}{\sqrt{2}v_u}
	  \end{pmatrix}.
\end{equation}
\end{widetext}
From Eq.~(\ref{CP-odd}), it is straightforward to show that the expected two Goldstone bosons are obtained.  For the CP-even neutral scalars, the mass matrix is:
\begin{equation}
\label{CP-even}
	{\cal M}_S^2	=
	\begin{pmatrix}
		S_{\nu}^2
		&
		S_{\nu H}^2
	\\
		\left(S_{\nu H}^{2}\right)^T
		&
		S_{H}^2
	\end{pmatrix},
\end{equation}
in which $S_{\nu}^2$, $S_{\nu H}^2$ and $S_H^2$ are given in the Appendix.
The mass matrix for the charged Higgses is given by 
\begin{equation}
\label{Charged}
	M_C^2 =
	\begin{pmatrix}
		C_e^2
		&
		C_{e H}^2
	\\
		\left(C_{e H}^2 \right)^T
		&
		C_{H}^2
	\end{pmatrix},
\end{equation}
in which the definitions of $C_{e}^2$, $C_{eH}^2$ and $C_{H}^2$ are also given in the Appendix. 

It is most illuminating to the study the scalar sector in the limit of zero neutrino 
masses $(v_L, \ Y_\nu, \ A_\nu \to 0)$, since the Higgs mass-squared matrices
greatly simplify.  In particular, the left-handed sneutrino decouples from the 
CP-even and CP-odd Higgses with a mass that is given by 
\begin{equation}
m_{\tilde \nu}^2 = M_{\tilde L}^2 - \frac{1}{8} g_{BL}^2 v_R^2 - \frac{1}{8} g_2^2 \left(v_u^2 - v_d^2\right).
\end{equation}
Imposing the experimental lower bound on the sneutrino masses results in a constraint 
on the relation between $M_{\tilde L}^2$ and the $B-L$ contribution.
The remaining $3\times 3$ CP-odd mass matrix has two zero eigenvalues; the corresponding states are the Goldstone bosons that are eaten by the $Z$ and $Z'$.  The nonzero eigenvalue corresponds to the CP-odd MSSM Higgs, $A^0$, which has the same expression for its mass as in the MSSM.  

The remaining $3\times 3$ CP-even mass matrix can be further studied in the limit of large
$\tan \beta$ and decoupling of the heavy CP-even MSSM Higgs.  This leaves a $2 \times 2$ mass
matrix with potentially large mixing between the right-handed 
sneutrino and the MSSM Higgs:
\begin{widetext}
\begin{equation}
	\mathcal{M}_S^2 \to
	\begin{pmatrix}
		\frac{1}{4} \left(g_{BL}^2 + g_R^2\right) v_R^2
		&
		-\frac{1}{4} g_R^2 v_u v_R
		\\
		-\frac{1}{4} g_R^2 v_u v_R
		&
		\frac{1}{4}\left(g_2^2 + g_R^2\right) v_u^2
	\end{pmatrix}.
\end{equation}
\end{widetext}
This matrix has the same trace and determinant as Eq.~(\ref{Z.Mass.Matrix}), indicating that
at zeroth order, the masses in this limit are given by
\begin{align}
	m^2_{\text{Re}\tilde \nu^C} & = M_{Z'}^2 \sim \frac{1}{4} \left(g_R^2 + g_{BL}^2\right) v_R^2,
	\\
	m^2_h & = M_Z^2 \sim \frac{1}{4} \left(g_1^2 + g_2^2\right) v^2.
\end{align}
In the limit of vanishing neutrino masses, the charged Higgs mass matrix of Eq.~(\ref{Charged}) decouples into a block diagonal form, in which the upper $2\times 2$ block resembles the MSSM slepton mass matrix and the lower $2\times 2$ block is identical to the MSSM charged Higgs mass matrix.  Assuming small mixing in the slepton sector, the slepton masses are given by
\begin{align}
	m_{\tilde e_L}^2 & = M_{\tilde L}^2 -\frac{1}{8} g_{BL}^2 v_R^2
				 + \frac{1}{8} g_2^2 \left(v_u^2 - v_d^2\right),
	\\
	m_{\tilde e_R}^2 & = M_{\tilde E^C}^2 + \frac{1}{8} g_{BL}^2 v_R^2
				 - \frac{1}{8} g_R^2 \left(v_R^2 + v_d^2 - v_u^2\right).
\end{align}
The leading order corrections to the scalar masses typically scale as $Y_\nu^2 v_R^2$.  
Such contributions are quite negligible as expected, since they are suppressed 
by the neutrino mass parameters.

As in any theory with R-parity violation, mixing between the charged 
leptons and the charginos of the MSSM will occur in the charged fermion sector, 
$\left(e^c, \ \tilde W_L^+, \ \tilde H_u^+ \right)$ and 
$\left( \ e, \ \tilde W_L^-, \ \tilde H_d^-\right)$. In this basis, 
the mass matrix is given by 
\begin{equation}
	{\cal M}_{\tilde C} =
	\begin{pmatrix}
			-\frac{1}{\sqrt{2}} Y_e v_d
		&
			0
		&
			\frac{1}{\sqrt{2}} Y_e v_L
	\\
			\frac{1}{\sqrt{2}} g_2 v_L
		&
			M_2
		&
			\frac{1}{\sqrt{2}} g_2 v_d
	\\
			-\frac{1}{\sqrt{2}} Y_\nu v_R
		&
			\frac{1}{\sqrt{2}} g_2 v_u
		&
			\mu
	\end{pmatrix}.
\end{equation}
Since the mixing between the charginos and the charged leptons of the MSSM is proportional to $v_L$ and $Y_\nu$,  small corrections to the charged lepton masses can be generated once the charginos are integrated out. However, this contribution is always small once the neutrino constraints are imposed.

The squark mass matrices in this theory differ from those obtained in the MSSM due to new D term contributions, as follows:
\begin{widetext}
\begin{eqnarray}
	{\cal M}_{\tilde u}^2 &=& 
	\begin{pmatrix}
		M_{\tilde Q}^2 + m_u^2 + \frac{1}{24} \ g_{BL}^2 \ v_R^2 
			- \frac{1}{8} \ g_2^2 \left(v_u^2 - v_d^2\right)
		&
		A_u \ v_u - y_u \ \mu \ v_d
		\\
		A_u \ v_u - y_u \ \mu \ v_d
		&
		M_{\tilde u^C}^2 + m_u^2 - \frac{1}{24} \ g_{BL}^2 \ v_R^2 
			+ \frac{1}{8} \ g_R^2 \left(v_R^2 + v_d^2 - v_u^2\right)
	\end{pmatrix},
	\nonumber \\ 
	\\
	{\cal M}_{\tilde d}^2 &=& 
	\begin{pmatrix}
		M_{\tilde Q}^2 + m_d^2 + \frac{1}{24} \ g_{BL}^2 \ v_R^2 
			+ \frac{1}{8} \ g_2^2 \left(v_u^2 - v_d^2\right)
		&
		A_d \ v_d - y_d \ \mu \ v_u
		\\
		A_d \ v_d - y_d \ \mu \ v_u
		&
		M_{\tilde d^C}^2 + m_d^2 - \frac{1}{24} \ g_{BL}^2 \ v_R^2 
			- \frac{1}{8} \ g_R^2 \left(v_R^2 + v_d^2 - v_u^2\right)
	\end{pmatrix},
	\nonumber \\
\end{eqnarray}
\end{widetext}
in which $m_u$ and $m_d$ are the up-type and down-type quark masses, respectively.  
Let us assume for simplicity that the LR mixings in the squark sector is small and thus can be neglected. The diagonal (LL and RR) terms must then be positive definite to avoid tachyonic squarks, as well as satisfy the experimental bounds.
Therefore $M_{\tilde d^C}^2 \ >  \  -m_d^2 + \frac{1}{24} \ g_{BL}^2 \ v_R^2 
			+ \frac{1}{8} \ g_R^2 \left(v_R^2 + v_d^2 - v_u^2\right)$ 
and $M_{\tilde u^C}^2 \ >  \ - m_u^2 + \frac{1}{24} \ g_{BL}^2 \ v_R^2 
			- \frac{1}{8} \ g_R^2 \left(v_R^2 + v_d^2 - v_u^2\right)$, in order to have a realistic
spectrum.
\section{Distinguishing between different SRpV scenarios}
\label{section4}
The collider phenomenology of this model at the LHC has distinctive features that are associated with the gauged $B-L$ at TeV energies and R-parity violating signatures.  The possibility of testing this class of models of spontaneous R-parity violation via the multilepton channels with different leptons, $eeee$, 
$e\mu \mu \mu$, $ee \mu \mu$ and others, from the sneutrinos decays $\tilde{\nu} \to e^+_i e^-_j$ has been discussed in \cite{Sogee3}. Within this class of models,  the sneutrinos can be pair produced through the  $Z^{'}$ gauge boson that present in these models; 
see~\cite{LTWang} for a detailed study of this production mechanism at the LHC.  

The models of spontaneous R-parity violation proposed here and in~\cite{Sogee2,Sogee3}  that involve a gauged $B-L$ symmetry have many similar phenomenological features.  Recall that these other scenarios are a minimal extension of the MSSM with $U(1)_{B-L}$ \cite{Sogee2}, and an extended version in which $U(1)_{B-L}$ is augmented by an admixture of $U(1)_Y$ \cite{Sogee3}.  Given the challenges of predicting the detailed decay patterns of the  neutralinos, $\tilde{\chi^0} \to W e, Z \nu$ and the other superpartner decays,  we believe 
that the easiest way to distinguish between these models (at least as a first step) is through the  properties of 
their $Z^{'}$ gauge bosons. In Table~\ref{Zprime couplings}, we have displayed the couplings of the $Z^{'}$ gauge boson to fermions in each scenario.  Note that these three scenarios are simple, anomaly-free models  with minimal particle content in which spontaneous R-parity breaking can be achieved and the properties of the $U(1)^\prime$ gauge boson are prescribed.  Hence, if a $Z^\prime$ gauge boson with some component of $U(1)_{B-L}$ is discovered at the LHC, the comparison of its properties from the pure $B-L$ case can in principle reveal if any of these scenarios for  spontaneous R-parity violation might be relevant (or other possibilities that we have not yet studied). Of course, we are considering the ideal case where SUSY with R-parity violating interactions is discovered at the LHC.  For further studies of how to characterize the couplings of a $Z^{'}$ at the LHC using leptonic channels as well as additional channels involving top and bottom quarks, we refer the reader to~\cite{Petriello,Godfrey}.
\begin{table}
\begin{center}
\begin{tabular}[t]{|c|c|}
  \hline
{\bf Models} & $ Z^{'} \bar{f} f $  Couplings\\
 \hline
 $U(1)_Y \bigotimes U(1)_{B-L}$ & $ g_{BL} (B-L) / 2$ \\
 \hline
 $U(1)_Y \bigotimes U(1)_X$ & $g_X \left(a \ Y + b \left(B-L\right)\right)/2$\\
 \hline
 $U(1)_{I^R_3} \bigotimes U(1)_{B-L}$ & $\sqrt{g_R^2 + g_{BL}^2} \
		\left[
			\cos \theta_R^2 \ Y(f)
			- I_{3R}(f)
		\right]$
		\\
 \hline
\end{tabular}
\end{center}
\caption{The $Z^{'}$ couplings to the fermions in the SRpV models involving a gauged $U(1)_{B-L}$ broken at the TeV scale studied here and in \cite{Sogee2,Sogee3}. In the above, $X= a Y + b (B-L)$.} 
\label{Zprime couplings}
\end{table}

We now discuss the main differences between the mass spectrum 
of the sleptons in each SRpV model. In Tables ~\ref{Model-I},
~\ref{Model-II} and~\ref{Model-III}, we show the mass spectrum in the limit $v_L \to 0$ for simplicity. Within a given SUSY breaking scenario that predicts $M_{\tilde L}^2$ and $M_{\tilde E^C}^2$, the main features of the spectrum in each case can be determined once values for $v_R$, $g_{BL}$, $g_X$, and $\tan \beta$ are assumed.  Hence, if SUSY is discovered at the LHC, measurements of the detailed properties of the slepton sector can indicate whether the MSSM or any of these extended models is suggested by the data.
\begin{table}[tb]
\begin{center}
\begin{tabular}[t]{|c|c|}
  \hline
  \hline
{\bf Slepton Masses} & $U(1)_Y \bigotimes U(1)_{B-L}$ \\
 \hline
 $m_{{\tilde e}_L}^2$ & $M_{\tilde L}^2 - \frac{1}{8} g_{BL}^2 v_R^2 + \frac{1}{8} (g_2^2 - g_1^2) (v_u^2 - v_d^2)$ \\
 \hline
 $m_{\tilde{e}_R}^2$  & $M_{\tilde E^C}^2 + \frac{1}{8} g_{BL}^2 v_R^2 + \frac{1}{4} g_1^2 (v_u^2 - v_d^2)$ \\
 \hline
 $m_{\tilde{\nu}_L}^2$ & $M_{\tilde L}^2 - \frac{1}{8} g_{BL}^2 v_R^2 - \frac{1}{8} (g_1^2 + g_2^2) (v_u^2 - v_d^2)$ \\
 \hline
 $m_{\tilde{\nu}^C}^2$ & $M_{Z_{B-L}}^2$ \\
  \hline
  \hline
\end{tabular}
\end{center}
\caption{The minimal $B-L$ model.} 
\label{Model-I}
\end{table}

\begin{table}[h]
\begin{center}
\begin{tabular}[t]{|c|c|}
  \hline
  \hline
{\bf Slepton Masses} & $U(1)_Y \bigotimes U(1)_{X}$ \\
 \hline
 $m_{{\tilde e}_L}^2$ & $M_{\tilde L}^2 + \frac{1}{32} g_{X}^2 ( - \frac{5}{4} v_R^2 + v_u^2 - v_d^2 ) 
 + \frac{1}{8} (g_2^2 - g_1^2) (v_u^2 - v_d^2)$ \\
 \hline
 $m_{\tilde{e}_R}^2$  & $M_{\tilde E^C}^2 
 + \frac{3}{32} g_{X}^2 ( - \frac{5}{4}v_R^2 + v_u^2 - v_d^2 ) + \frac{1}{4} g_1^2 (v_u^2 - v_d^2)$ \\
 \hline
 $m_{\tilde{\nu}_L}^2$ & $M_{\tilde L}^2 
 + \frac{1}{32} g_{X}^2 ( - \frac{5}{4} v_R^2 + v_u^2 - v_d^2) - \frac{1}{8} (g_1^2 + g_2^2) (v_u^2 - v_d^2)$ \\
 \hline
 $m_{\tilde{\nu}^C}^2$ & $M_{Z_X}^2$ \\
  \hline
  \hline
\end{tabular}
\end{center}
\caption{The $U(1)_Y \bigotimes U(1)_{X}$ model, in which $X= a Y + b (B-L)$  with $a=1$ and $b=-5/4$.} 
\label{Model-II}
\end{table}

\begin{table}[h!]
\begin{center}
\begin{tabular}[t]{|c|c|}
  \hline
  \hline
{\bf Slepton Masses} & $U(1)_{I^3_R} \bigotimes U(1)_{B-L}$ \\
 \hline
 $m_{{\tilde e}_L}^2$ & $M_{\tilde L}^2 - \frac{1}{8} g_{BL}^2 v_R^2 + \frac{1}{8} g_2^2  (v_u^2 - v_d^2)$ \\
 \hline
 $m_{\tilde{e}_R}^2$  & $M_{\tilde E^C}^2 + \frac{1}{8} g_{BL}^2 v_R^2 - \frac{1}{8} g_R^2 (v_R^2 + v_d^2 - v_u^2)$ \\
 \hline
 $m_{\tilde{\nu}_L}^2$ & $M_{\tilde L}^2 - \frac{1}{8} g_{BL}^2 v_R^2 - \frac{1}{8} g_2^2 (v_u^2 - v_d^2)$ \\
 \hline
 $m_{\tilde{\nu}^C}^2$ & $M_{Z_{LR}}^2$ \\
  \hline
  \hline
\end{tabular}
\end{center}
\caption{The LR-inspired model described here, in which the SM hypercharge is $g_1=g_R g_{BL}/\sqrt{ g_{BL}^2 + g_R^2}$.} 
\label{Model-III}
\end{table}

\newpage
\section{Summary and Discussion}
\label{section5}
In this paper, we have discussed spontaneous R-parity violation in the context of a simple 
extension of the minimal supersymmetric Standard Model where the Abelian sector of the theory is composed of $B-L$ and right-handed isospin.   This model can be obtained as a low energy limit of left-right symmetric scenarios in which the scale of left-right symmetry breaking is much higher than the scale of R-parity breaking.   We have analyzed the gauge symmetry breaking and low energy mass spectrum of the theory, including the generation of neutrino masses in this scenario.  The nature of the resulting R-parity violating interactions and their phenomenological implications are also presented.  We also described how to distinguish between this and other simple scenarios which include a gauged $U(1)_{B-L}$ as part of the Abelian gauge sector of the theory at the LHC via their $Z^\prime$ phenomenology. 

Within models with a gauged $U(1)_{B-L}$, if a minimal particle content is assumed then R-parity will be spontaneously broken.  Indeed, in this class of models, R-parity is only an exact symmetry if the Higgs sector is extended to include additional SM singlets which have even $B-L$ quantum numbers. Only experiments will tell us if R-parity is an exact symmetry at low energies, but in our view, the possibility of spontaneous R-parity violation as investigated in this paper is quite appealing and provides phenomenological signatures that should be considered in searches for supersymmetry at TeV energies. 

The simple and economical model studied in this work not only has direct connections with grand unification, but as with other scenarios with a gauged $B-L$ symmetry that is broken at TeV energies, it provides a simple and calculable scenario in which to study R-parity violation at the LHC.  Such models of spontaneous R-parity violation avoid the standard concerns of supersymmetric theories associated with rapid proton decay, and can provide in the gravitino a viable dark matter candidate.   Furthermore, this class of models also yield simple, anomaly free $U(1)^\prime$ gauge extensions of the MSSM with distinctive couplings of the $Z^\prime$ to the fermions in the theory.  The $Z^\prime$ phenomenology of such simple $U(1)^\prime$ scenarios  is certainly worthy of further study.  Given these attractive features, such well-motivated alternatives to R-parity conserving models such as the MSSM (and many of its extensions) warrant further attention in this era of unprecedented exploration of TeV scale physics at the LHC.


\subsection*{Acknowledgments}
P.~F.~P. is supported in part by the U.S. Department of Energy contract No. DE-FG02-08ER41531 and the Wisconsin Alumni Research Foundation. L.~E. and S.~S. are supported in
part by the U.S. Department of Energy under grant No. DE-FG02-95ER40896,
and the Wisconsin Alumni Research Foundation.\\
\appendix
\section{General Scalar Potential and Mass Matrices}
Here we discuss the properties of the scalar potential, taking into account the possibility of vacuum expectation values for all three generations of the sneutrinos. 
Neglecting CP violating effects and defining $<H_u^0>=v_u/\sqrt{2}$, $<H_d^0>=v_d/\sqrt{2}$, $<\tilde{\nu}^i>=v_L^i/\sqrt{2}$, and $<(\tilde{\nu}^C)^i>=v_R^i/\sqrt{2}$,  the scalar potential is
\begin{eqnarray}
V &=& \frac{1}{4} \left( v_L^i \ Y_\nu^{ij} \ v_R^j \right)^2 \ + \ \frac{\mu^2}{2} (v_u^2 \ + \ v_d^2) 
 - \frac{\mu}{\sqrt{2}}  v_L^i \ Y_\nu^{ij} \ v_R^j \ v_d \ - \  \frac{1}{4}  v_u^2 \left(  Y_\nu^{ij} \ v_R^j \ Y_\nu^{ik} \ v_R^k 
 \ + \  v_L^i \ Y_\nu^{ij} \ v_L^k  Y_\nu^{kj}\right)
 \nonumber \\
& + & \frac{g_2^2}{32} (v_L^i v_L^i \ - \ v_u^2 \ + \ v_d^2)^2 \ + \ \frac{g_R^2}{32} ( v_R^i v_R^i \ - \ v_L^i v_L^i )^2 
\ + \ \frac{1}{2} v_R^i (M_{\tilde{N}^C}^2)_{ij} v_R^j   + \frac{1}{2} v_L^i (M_{\tilde L}^2)_{ij} v_L^j  
+\ \frac{1}{2} m_{H_u}^2 v_u^2 
\nonumber \\
& + & \frac{1}{2} m_{H_d^2} v_d^2  \ + \ \frac{1}{2}
\left(  \frac{1}{\sqrt 2} v_L^i \ A_\nu^{ij} \ v_R^j v_u \ - \  B \mu  v_u v_d  \ + \ \rm{h.c.}\right). 
\end{eqnarray}
In the case in which only one generation of sneutrinos acquire vacuum expectation values and in the limit that the Yukawa coupling and the trilinear term are flavor diagonal, the results of Section II are reproduced.

In the $\sqrt{2} \real \left(\tilde \nu, \tilde \nu^c, H_d^0, H_u^0 \right)$ basis
for the CP-even scalars and the $\left(\tilde e^*, \tilde e^c, H_d^{-*}, H_u^+ \right)$ basis for 
the charged scalars, the mass matrices for these two sectors are given by Eq.~(\ref{CP-even}) and Eq.~(\ref{Charged}), 
in which 
\begin{widetext}
\begin{align}
	S_\nu^2	\equiv	&
	\begin{pmatrix}
		\frac{1}{4} \left(g_2^2 + g_{BL}^2 \right) v_L^2 
		+ \frac{v_R}{v_L} B_\nu
		&
		-\frac{1}{4} g_{BL}^2 v_L v_R + \left(Y_\nu \right)^2 v_L v_R - B_\nu
	\\
		-\frac{1}{4} g_{BL}^2 v_L v_R + \left(Y_\nu \right)^2 v_L v_R - B_\nu
		&
		\frac{1}{4} \left(g_{BL}^2 + g_R^2\right) v_R^2 + \frac{v_L}{v_R} B_\nu
	\end{pmatrix},
\\
\nonumber
\\
	S_{\nu H}^2 \equiv	&
	\begin{pmatrix}
		\frac{1}{4} g_2^2 v_d v_L  
		- \frac{1}{\sqrt{2}} Y_\nu \mu \ v_R
		&
		-\frac{1}{4} g_2^2 v_L v_u + \left(Y_\nu \right)^2 v_L v_u
		+ \frac{1}{\sqrt{2}} A_\nu v_R
	\\
		\frac{1}{4} g_{R}^2 v_d v_R-\frac{1}{\sqrt{2}} Y_\nu \mu \ v_L
		&
		-\frac{1}{4} g_{R}^2 v_u v_R+ \left(Y_\nu \right)^2 v_u v_R
		+\frac{1}{\sqrt{2}} A_\nu v_L
	\end{pmatrix},
\\
\nonumber
\\
	S_{H}^2	\equiv &
	\begin{pmatrix}
		\frac{1}{4} \left(g_2^2 + g_R^2\right)v_d^2 + \frac{v_u}{v_d} B\mu  
		+ \frac{Y_\nu \mu \ v_L v_R}{\sqrt{2} \ v_d}
		&
		 - \frac{1}{4} \left(g_2^2 + g_R^2\right) v_u v_d - B\mu
	\\
		 - \frac{1}{4} \left(g_2^2 + g_R^2\right) v_u v_d - B\mu
		&
		\frac{1}{4} \left(g_2^2  + g_R^2\right) v_u^2 + \frac{v_d}{v_u} B\mu
		 - \frac{A_\nu v_L v_R}{\sqrt{2} \ v_u}
	\end{pmatrix}.	
\end{align}
\begin{align}
C_e^2 \equiv &
	\begin{pmatrix}
		C_{11}^2
		&
		B_e
	\\
		B_e
		&
		C_{22}^2
	\end{pmatrix},
\end{align}
\begin{align}
	C_{e H}^2 \equiv
	&
	\frac{1}{2}\begin{pmatrix}
		\frac{1}{2} g_2^2 \ v_d v_L - Y_e^2  v_d v_L 
		-  Y_\nu \mu \ v_R
		&
		\frac{1}{2} g_2^2 v_L v_u -  Y_\nu^2 v_u v_L 
		- \sqrt{2} A_\nu v_R
	\\
		 Y_e Y_\nu v_u v_R + \sqrt{2} A_e v_L
		&
		Y_e Y_\nu v_d v_R + \sqrt{2} Y_e \mu \ v_L
	\end{pmatrix},
\end{align}
\begin{align}
\nonumber
\\
	C_{H}^2 \equiv
	&
	\begin{pmatrix}
		\frac{1}{4} g_2^2\left(v_u^2-v_L^2 \right) + B \mu \frac{v_u}{v_d} 
		+ \frac{1}{2} Y_e^2 v_L^2 + \frac{Y_\nu \mu \ v_R v_L}{\sqrt{2} \ v_d}
		&
		B\mu + \frac{1}{4} g_2^2 v_u v_d
	\\
		B\mu + \frac{1}{4} g_2^2 v_u v_d
		&
		\frac{1}{4} g_2^2 \left(v_d^2 + v_L^2 \right) + \frac{v_d}{v_u}B\mu 
		- \frac{1}{2} \left(Y_\nu \right)^2 v_L^2 - \frac{A_\nu v_L v_R}{\sqrt{2} \ v_u}
	\end{pmatrix},
\end{align}
with
\begin{eqnarray}
C_{11}^2 & = & \frac{1}{4} g_2^2 \left(v_u^2 - v_d^2 \right) + 
	\frac{1}{2}Y_e^2 v_d^2 - \frac{1}{2} {Y_\nu}^2 v_u^2 + \frac{v_R}{v_L} B_\nu,\\
C_{22}^2 &=& M_{\tilde E^c}^2 -\frac{1}{8} g_R^2 \left(v_R^2 + v_d^2 - v_u^2\right) 
			+ \frac{1}{8} g_{BL}^2 \left(v_R^2 - v_L^2\right)
			+ \frac{1}{2} Y_e^2 \left( v_d^2 + v_L^2 \right),
\end{eqnarray}
\end{widetext}
where
	$B_\nu  \equiv \frac{1}{\sqrt{2}} \left(Y_\nu \mu v_d - A_\nu v_u \right)$, 
	\text{ and }
	$B_e  \equiv \frac{1}{\sqrt{2}} \left(Y_e \mu \ v_u - A_e v_d\right)$.

\section{Fayet-Iliopoulos Terms}
We now describe the case in which the potential includes Fayet-Iliopoulos D terms:
$V_{FI}= \xi_R D_R + \xi_{BL} D_{BL}$. In this case the minimization 
conditions take the form
\begin{align}
	v_R & = \sqrt
				{
					\frac{-8 M_{\tilde{N}^C}^2 
					- 4\left(g_R \xi_R - g_{BL} \xi_{BL}\right)
					+ g_R^2 \left(v_u^2 - v_d^2\right)}
					{g_R^2 + g_{BL}^2}
				},
	\\
	v_L & = \frac{v_R B_\nu}
			{
				M_{\tilde L}^2 
				+ \frac{1}{2} g_{BL} \xi_{BL}
				- \frac{1}{8} g_{BL}^2 v_R^2 
				- \frac{1}{8} g_2^2 \left(v_u^2 - v_d^2\right)
			},
	\\
	\mu^2 & = - \frac{1}{8} \left(g_2^2 + g_R^2\right) \left(v_u^2 + v_d^2\right)
			+ \frac{M_{H_u}^2 \tan^2 \beta  - M_{H_d}^2}{1-\tan^2 \beta},
	\\
	B\mu & =  2 \mu^2 + m_{H_d}^2 + m_{H_u}^2,
\end{align}
in which 
\begin{align}
	M_{H_u}^2 = m_{H_u}^2 - \frac{1}{2} g_R \xi_R - \frac{1}{8} g_R^2 v_R^2,  
	M_{H_d}^2 = m_{H_d}^2 + \frac{1}{2} g_R \xi_R + \frac{1}{8} g_R^2 v_R^2,
	B_\nu = \frac{1}{\sqrt{2}} \left(Y_\nu \mu v_d - A_\nu v_u \right).
\end{align}
If the $B-L$ FI-term, $\xi_{BL}$, is large and positive, note that we can have a consistent mechanism for spontaneous R-parity violation even with positive soft mass-squared terms for the sneutrinos. 
In this case, 
\begin{eqnarray}
C_{22}^2 = 
	M_{\tilde E^c}^2 
	- \frac{1}{2} \left(g_R \xi_R + g_{BL} \xi_{BL}\right)
	-\frac{1}{8} g_R^2 \left(v_R^2 + v_d^2 - v_u^2\right) 
	\ + \ \frac{1}{8} g_{BL}^2 \left(v_R^2 - v_L^2\right)
	+\frac{Y_e^2}{2} ( v_d^2 + v_L^2),
\end{eqnarray}
and the squark masses (assuming negligible mixing) read as
\begin{eqnarray}
	m_{\tilde u_L}^2 &
		= &M_{\tilde Q}^2
		- \frac{1}{6} g_{BL} \xi_{BL}
		+ \frac{1}{24} \ g_{BL}^2 \ v_R^2
		\ + \ \frac{1}{8} \ g_2^2 \left(v_u^2 - v_d^2\right)
	\\
	m_{\tilde d_L}^2 &
		= &M_{\tilde Q}^2
		- \frac{1}{6} g_{BL} \xi_{BL}
		+ \frac{1}{24} \ g_{BL}^2 \ v_R^2 
		\ - \ \frac{1}{8} \ g_2^2 \left(v_u^2 - v_d^2\right)
	\\
	m_{\tilde u_R}^2 &
		= &M_{\tilde U^C}^2
		+ \frac{1}{6} g_{BL} \xi_{BL}
		+ \frac{1}{2} g_{R} \xi_{BL}
		- \frac{1}{24} \ g_{BL}^2 v_R^2 
		\ + \ \frac{1}{8} \ g_R^2 \left(v_R^2 + v_d^2 - v_u^2\right)
	\\
	m_{\tilde d_R}^2 &
		=& M_{\tilde D^C}^2
		+ \frac{1}{6} g_{BL} \xi_{BL}
		- \frac{1}{2} g_{R} \xi_{BL}
		- \frac{1}{24} \ g_{BL}^2 v_R^2 
		\ - \ \frac{1}{8} \ g_R^2 \left(v_R^2 + v_d^2 - v_u^2\right).
\end{eqnarray}
Again, once the bounds on the superpartner masses are imposed 
interesting correlations can be obtained. Further details of the phenomenology of this model will be investigated in a future publication.
%

\end{document}